\documentclass[aps,twocolumn,prx,floatfix,showpacs,citeautoscript,superscriptaddress]{revtex4-1}

\usepackage{amssymb}
\usepackage{amsmath}
\usepackage{graphicx}
\usepackage{bm}
\usepackage{times}
\usepackage{color}
\usepackage[dvipsnames,svgnames,table]{xcolor}
\usepackage{hyperref}
\usepackage{url}
\usepackage{fancyhdr}
\usepackage{float}
\usepackage{enumitem}
\usepackage[caption=false]{subfig}
\usepackage{mathrsfs}
\usepackage[scr=boondox,  
            cal=esstix]   
           {mathalpha}
\hypersetup{
pdfstartview={FitH},
colorlinks=true,    
linkcolor=NavyBlue, 
citecolor=Blue,   
filecolor=NavyBlue, 
urlcolor=NavyBlue   
}

\begin{document}

\title{Fluctuation amplification engineering in multimode Raman-cavity systems}

\author{H. P. {Ojeda Collado}}
\email{hojedaco@physnet.uni-hamburg.de }
\affiliation{Center for Optical Quantum Technologies and Institute for Quantum Physics, University of Hamburg, 22761 Hamburg, Germany}
\affiliation{The Hamburg Center for Ultrafast Imaging, Luruper Chaussee 149, 22761 Hamburg, Germany}
\author{Ludwig Mathey}
\affiliation{Center for Optical Quantum Technologies and Institute for Quantum Physics, University of Hamburg, 22761 Hamburg, Germany}
\affiliation{The Hamburg Center for Ultrafast Imaging, Luruper Chaussee 149, 22761 Hamburg, Germany}

\begin{abstract}

Parametric amplification is a key ingredient of a wide range of phenomena, from the classical to the quantum domain. Although such phenomena have been demonstrated in non-equilibrium settings, their use for fluctuation engineering has been put forth in Raman-cavity hybrids only recently. In this work, we generalize fluctuation engineering to a multi-mode scenario in which multiple Raman-active modes interact nonlinearly with multiple cavity modes. We demonstrate the emergence of resonant and non-resonant collective fluctuations that can be non-reciprocally controlled by engineering the band dispersion of photons and phonons. As an example we show how Raman fluctuations can be selectively attenuated by tuning the photonic bandgap or even nonresonantly amplified, in marked contrast to the single-mode scenario. We also identify a regime in which the amplification of cavity fluctuations in a specific mode is boosted, surpassing a $\sqrt{N}$ scaling with increasing number of $N$ Raman and cavity modes.
Our study reveals the key role of multi-mode interactions on fluctuations in nonlinear cavity-matter hybrids. Noise engineering through different photon and phonon dispersions, as demonstrated here, could be leveraged for the design of novel quantum sensing platforms and advanced spectroscopy in the THz regime.

\end{abstract}

\date{\today}

\maketitle

\section{Introduction}
Controlling many-body systems via laser light has generated novel transient phenomena and nonequilibrium states of matter. Notable examples include the control of ferroelectricity~\cite{Nova19}, magnetism~\cite{disa_photo-induced_2023,siegrist_light-wave_2019,Samuel_21,Emil_20}, charge density waves~\cite{kogar_light-induced_2020,dolgirev_self-similar_2020} and superconductivity~\cite{mitrano_possible_2016,budden_evidence_2021}. In such nonequilibrium scenarios, it is even possible to create new phases of matter without an equilibrium counterpart, such as time crystals, which have been theoretically proposed and experimentally observed on various platforms~\cite{Zhang2017,Choi2017,Else2020,Kessler2021,Kongkhambut2021,Yao17,Pizzi,HP_2021,HP_2023,Taheri2022,Zaletel2023,Carraro,Greilich}.

A promising alternative to optical control via laser light is the utilization of zero-point-energy light fluctuations through cavities, i.e. cavity control. This equilibrium approach offers the possibility of controlling the properties of the materials while preventing unwanted heating effects that could be generated by high-intensity lasers. The control of quantum materials using cavities has been demonstrated experimentally in various contexts, including the manipulation of transport~\cite{orgiu_conductivity_2015}, the control of superconducting properties~\cite{thomas_exploring_2019}, chemical reactions~\cite{thomas_tilting_2019,nagarajan_chemistry_2021,Sclaefer_19}, magnetism~\cite{thomas_large_2021} and topology~\cite{appugliese_breakdown_2022}. Furthermore, the use of cavities to enhance spectroscopic signals~\cite{Fainstein95,Gagliardi13,Melanie16,Sulzer22} shows that controlling light fluctuations in a cavity can be exploited in the design of novel quantum sensing techniques. 

A typical regime of light- or cavity-induced physics is the one in which the interaction between light and matter is stronger than the decay rates associated with the different dissipative channels. In this regime novel cavity-matter hybrids arise, such as polaritons~\cite{Polaritonpanorama}. One relevant task to advance the field of cavity-matter hybrids at equilibrium is to find novel mechanisms of control of matter and light properties. Of particular interest are those hybrids in which the fluctuations can be parametrically controlled. An example of this is the recent theoretical demonstration of parametric amplification of vacuum fluctuations and phonon localization in single-mode Raman-cavity hybrids~\cite{HP24}.

\begin{figure}[!tb]
\hspace*{-0.0cm} 
\includegraphics[width=0.35\textwidth]{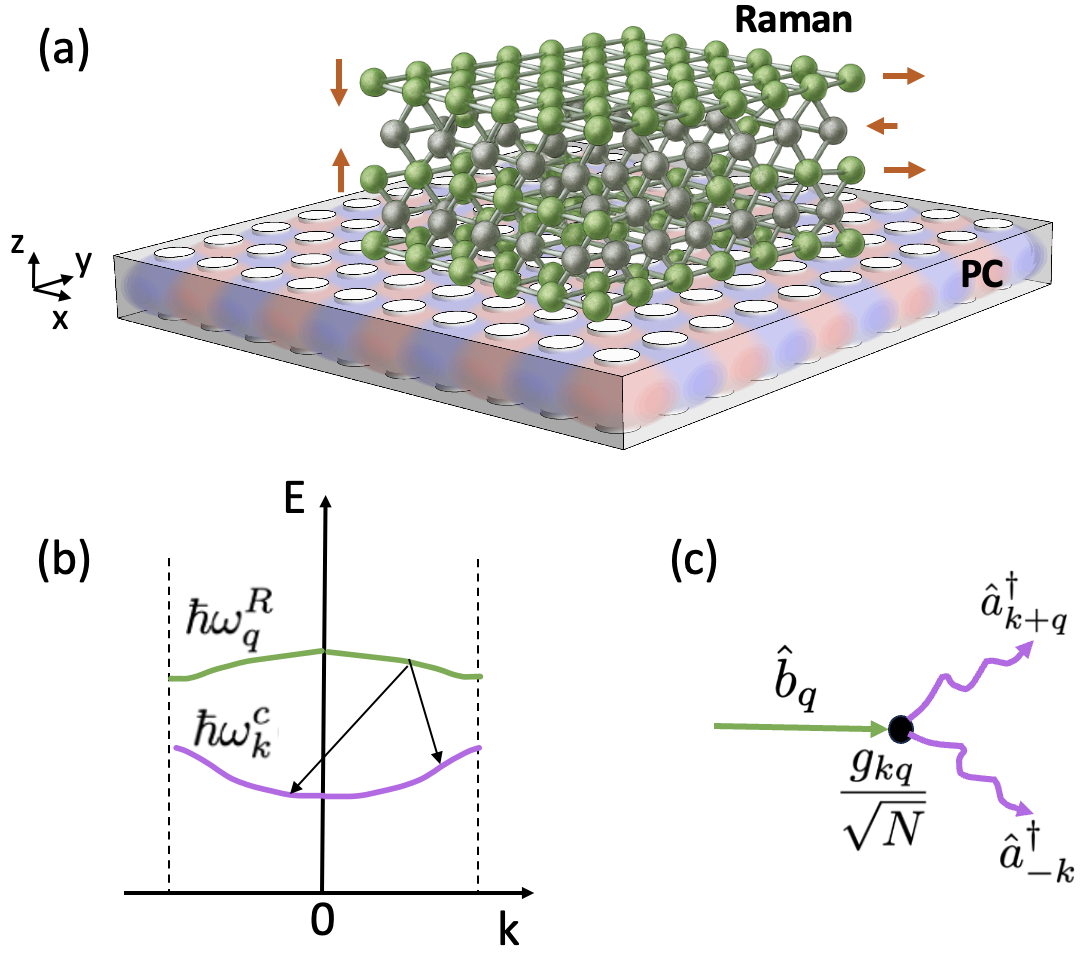}
\caption{Raman-active modes coupled to light fluctuations of a photonic crystal. (a) Sketch of a Raman-active material on top of a photonic crystal (PC). The red arrows represent an example of THz Raman-active phonon oscillations. Specifically, breathing (arrows to the left) and shearing (arrows to the right) modes that couple to light fluctuations of PC. (b) Schematic of Raman phonon dispersion (green) and the photonic crystal dispersion (magenta). 
(c) Vertex interaction of strength $g_{kq}/\sqrt{N}$, where one Raman mode converts to two photonic modes by conserving momentum. This process is also illustrated by arrows in (b). $N$ is the number of unit cells. 
} 
\label{fig:cartoon1}
\end{figure} 

In this paper, we generalize this previous work of a single cavity mode coupled to a single Raman mode to include multiple Raman-active modes that couple to multiple cavity modes (see Fig.~\ref{fig:cartoon1}).
Starting from a macroscopic theory of inelastic Raman scattering, we derive a many-body interacting model and characterize the new ground state of the hybrid system in the presence of dissipation. We treat the problem using the open-truncated Wigner approximation (TWA) and explore different phonon and photon dispersions. Depending on those dispersions, we find strikingly different results. For the case of photon and phonon flat bands, we demonstrate collective light fluctuations amplification and Raman fluctuations attenuation for each mode that scales as $\sqrt{N}$ with the number of modes $N$. We show two representative examples in which either the photon or phonon band is dispersive while the second band remains flat. In such cases, we demonstrate (a) nonreciprocal cavity control of phonons, in which Raman mode fluctuations can be selectively reduced or even nonresonantly amplified by tuning the photonic band gap and (b) a further enhancement of light fluctuation amplification in certain modes that exceed the scaling $\sqrt{N}$.

We first focus on a Raman-active material that couple to a photonic crystal (Fig.~\ref{fig:cartoon1}) and later discuss different platforms where our model could be experimentally realized, including two-dimensional materials integrated with THz cavities and trapped ions as a phonon-based simulator of our model.

\section{Multimode Raman-cavity system} \label{Raman-Dicke_sec}
Unlike infrared phonon modes, Raman-active phonons do not carry a dipole moment. Therefore, the coupling of Raman modes to light is a second-order process.
Our starting point is the macroscopic theory of non-resonant inelastic Raman scattering, known as Placzek's approximation~\cite{Raman_scattering_Reference2,Raman_scattering_Reference5,Raman_scattering_Reference4}. Here, Raman phonons interact with light via a change in the electric polarizability. 
In this approach, we consider the Raman and light frequencies to be out of resonance with respect to the energy scales of dipole-active intermediate electronic states such as excitons.



We consider a Raman-active material that is coupled to a multimode photonic crystal, as depicted in Fig.~\ref{fig:cartoon1}. The electric field of light in the photonic crystal at position $\mathbf{r}$, $\mathbf{E}(\mathbf{r})$, induces a polarization in the Raman material which is given by $P_i(\mathbf{r})=\sum_{j}\chi_{ij} E_j(\mathbf{r})$, where $\chi_{ij}$ is the electric polarizability tensor, and $i$ and $j$ denote the three different spatial components; $x, y, z$. The energy stored in such polarizable material is
\begin{equation}
    H=\frac{\epsilon_0}{2} \int d^3r \sum_{i,j} E_i(\mathbf{r})
     \chi_{ij} E_j(\mathbf{r}).
\end{equation}
where $\epsilon_0$ is the vacuum permittivity. The electric polarizability tensor $\chi_{ij}$ is not a constant but depends on the Raman phonon displacements $Q(\mathbf{r})$ in the material. By considering small displacements around the equilibrium position of the atoms in the unit cell, we expand the polarizability to the first order in the Raman coordinate and obtain the Raman-light interaction in Placzek’s approximation as
\begin{equation}
H_{int}=\frac{\epsilon_0}{2} \int d^3\mathbf{r} \sum_{i,j} E_i(\mathbf{r}) \left( \frac{\partial \chi_{ij}}{\partial Q(\mathbf{r})} \right)_0Q(\mathbf{r}) E_j(\mathbf{r}). 
\label{Hint}
\end{equation}
Now we quantize the electromagnetic field in the photonic crystal, the Raman displacement, and their mutual interaction [Eq.(\ref{Hint})] by considering the geometry shown in Fig.~\ref{fig:cartoon1} (a). We consider a finite photonic crystal that is extended in the $xy$ plane and thin enough in the $z$-axis so that the dielectric constant is periodic in the $xy$ plane and constant along the $z$-axis. Therefore, the electric field operator is written as
\begin{multline}
  \mathbf{\hat{E}}(\mathbf{r}) = \sum_\mathbf{k} \sqrt{\frac{\hbar \omega_\mathbf{k}^c}{2 \epsilon_0 V_c}} \left[ \mathbf{e}_{\mathbf{k}{||}} \cos(k_z z)(\hat{a}_\mathbf{k} + \hat{a}_{-\mathbf{k}}^\dagger)\right] e^{i(k_x x + k_y y)}  \\
-i\sqrt{\frac{\hbar \omega_\mathbf{k}^c}{2 \epsilon_0 V_c}}\left[ \mathbf{e}_{\mathbf{k}_{\perp}} \sin(k_z z)(\hat{a}_\mathbf{k} - \hat{a}_{-\mathbf{k}}^\dagger) \right]e^{i(k_x x + k_y y)},   
\end{multline}
where
\begin{equation}
k_x = \frac{2n_x \pi}{L_x}, \quad k_y = \frac{2n_y \pi}{L_y}, \quad k_z = \frac{n_z \pi}{L_z},
\label{discrete}
\end{equation} $n_x,n_y$ are integers and $n_z=0$ due to the geometry of the system. $L_x$, $L_y$, $L_z$ are the spatial dimension of the photonic crystal, and $V_c$ is the effective volume. $\mathbf{e}_{\mathbf{k}{||}}$ denotes the part of the unit vector parallel to the surface, while $\mathbf{e}_{\mathbf{k}{\perp}}$ is the perpendicular component. $\omega_\mathbf{k}^c$ is the cavity photon dispersion. $\hat{a}_\mathbf{k}$ ($\hat{a}^\dagger_\mathbf{k}$) is the photonic operator that annihilates (creates) a photon with momentum $\mathbf{k}$.

Considering a generic phonon unit cell and the same boundary conditions as for the electric field, the Raman coordinate reads
\begin{multline}
 \hat{Q}(\mathbf{r}) = \sum_{\mathbf{q}} \sqrt{\frac{\hbar}{2 M \omega_{\mathbf{q}}^R N}} \left[ \hat{e}_{q_{||}} \cos(q_z z)(\hat{b}_{\mathbf{q}} + \hat{b}_{-\mathbf{q}}^\dagger)\right] e^{i(q_x x + q_y y)}  \\
-i\sqrt{\frac{\hbar}{2 M \omega_{\mathbf{q}}^R N}}\left[ \hat{e}_{q_{\perp}} \sin(q_z z)(\hat{b}_{\mathbf{q}} - \hat{b}_{-\mathbf{q}}^\dagger) \right]e^{i(q_x x + q_y y)}   
\end{multline}
where $\hat{b}_{\mathbf{q}}$ ($\hat{b}^\dagger_{\mathbf{q}}$) annihilates (creates) a phonon with momentum $\mathbf{q}$ with a similar discretization as in Eq.~(\ref{discrete}). In this work, we consider a single photonic band and a single Raman phonon band as illustrated in Fig.~\ref{fig:cartoon1} (b). For phonons, it corresponds to a single-atom unit cell where $M$ is the mass of the atom, $N$ is the number of unit cells and $\omega_\mathbf{q}^R$ is the Raman phonon dispersion. These considerations remain a good approximation in a multiband scenario as long as the separation between the bands is greater than the variation of the photonic band gap that we will consider in the following sections.

The integral over space in Eq.~(\ref{Hint}) converts to a sum of integrals over each individual unit cell, and the total Hamiltonian, including the free electromagnetic field and Raman field, can be written as
\begin{multline}
 H = \sum_\mathbf{k} \hbar\omega^c_{\mathbf{k}}(\hat{a}_{\mathbf{k}}^\dagger \hat{a}_{\mathbf{k}}+\frac{1}{2})+ \sum_{\mathbf{q}} \hbar\omega_{\mathbf{q}}^R(\hat{b}_{\mathbf{q}}^\dagger \hat{b}_{\mathbf{q}}+\frac{1}{2}) \\
 +\sum_\mathbf{\mathbf{k},\mathbf{q}} \frac{g_{\mathbf{k}\mathbf{q}}}{\sqrt{N}}(\hat{a}_{\mathbf{k}}+ \hat{a}_{-\mathbf{k}}^\dagger)(\hat{b}_{\mathbf{q}}+ \hat{b}_{-\mathbf{q}}^\dagger)(\hat{a}_{-\mathbf{k}-\mathbf{q}}+ \hat{a}_{\mathbf{k}+\mathbf{q}}^\dagger)
\end{multline}
Following Ref.~\cite{HP24}, we consider an additional photonic quartic term to this Hamiltonian to ensure the stability of the system. The final Hamiltonian reads
\begin{multline}
 H = \sum_\mathbf{k} \hbar\omega^c_{\mathbf{k}}(\hat{a}_{\mathbf{k}}^\dagger \hat{a}_{\mathbf{k}}+\frac{1}{2})+ \sum_{\mathbf{q}} \hbar\omega^R_{\mathbf{q}}(\hat{b}_{\mathbf{q}}^\dagger \hat{b}_{\mathbf{q}}+\frac{1}{2}) \\
 +\sum_{\mathbf{k},\mathbf{q}} \frac{g_{\mathbf{k}\mathbf{q}}}{\sqrt{N}}(\hat{a}_{\mathbf{k}}+ \hat{a}_{-\mathbf{k}}^\dagger)(\hat{b}_{\mathbf{q}}+ \hat{b}_{-\mathbf{q}}^\dagger)(\hat{a}_{-\mathbf{k}-\mathbf{q}}+ \hat{a}_{\mathbf{k}+\mathbf{q}}^\dagger)  \\
 +\frac{g_4}{N} \sum_{\mathbf{k},\mathbf{k^\prime},\mathbf{q}} (\hat{a}_{-\mathbf{k}}+ \hat{a}_{\mathbf{k}}^\dagger)(\hat{a}_{\mathbf{k}+\mathbf{q}}+ \hat{a}_{-\mathbf{k}-\mathbf{q}}^\dagger)  \\
\quad \times (\hat{a}_{\mathbf{k^\prime}}+ \hat{a}_{-\mathbf{k^\prime}}^\dagger)(\hat{a}_{\mathbf{-k^\prime}-\mathbf{q}}+ \hat{a}_{\mathbf{k^\prime}+\mathbf{q}}^\dagger).
    \label{HamiltonianMB}
\end{multline}
The Raman-light coupling is, in general, momentum dependent and is given by $g_\mathbf{\mathbf{k}\mathbf{q}}=\frac{\tilde{R}_\mathbf{kq}V_s\hbar}{4 V_c}\sqrt{\omega_\mathbf{k}^c\omega_{-\mathbf{k}-\mathbf{q}}^c}$ where $V_s$ is the volume of the Raman medium and
\begin{align}
    \tilde{R}_\mathbf{kq} =& \mathbf{e}^{\mathbf{k}}_{c} \cdot \sum_{\mu = 1}^{N_a}\sum_{i = {x,y,z}} \frac{\partial \chi}{\partial i(\mu)} \frac{e^\mathbf{q}_i(\mu)}{\sqrt{2M\omega_{\mathbf{q}}^R} } \cdot \mathbf{e}^{-\mathbf{k}-\mathbf{q}}_{c}
    \label{Ramantensor}
\end{align}
is the dimensionless Raman tensor. The first derivative of electric polarizability in  Eq.~(\ref{Ramantensor}) is taken over the three different spatial directions, which gives rise to three different matrices. $\mu$ is an index that runs over the $N_a$ atoms of the unit cell, $e^{\mathbf{q}}_{i}(\mu)$ is the $i$-th component of the Raman displacement eigenvectors of the dynamical matrix associated with the Raman mode $\mathbf{q}$ with eigenvalue $\omega_{\mathbf{q}}^R$. $e^{\mathbf{k}}_{c}$ is the unit vector associated with the polarization of the $\mathbf{k}$ mode of the cavity field.
\section{Model and Methodology}
To illustrate the collective fluctuation amplification phenomenon, we study the variation of Raman and cavity fluctuations resulting from the hybridization in the many-body scenario. We present different examples of phonon and photon dispersions, and, for simplicity, we consider that the coupling between the modes is constant and $\mathbf{k}$-independent, so $g_{\mathbf{kq}}\equiv g$. Specifically, we determine the modification of the variance of the Raman and cavity fields in each mode as the consequence of coupling $g$ as
\begin{equation}
 \delta V(Q_{\mathbf{k}})=\frac{V(Q_\mathbf{k})_g-V(Q_\mathbf{k})_0}{V(Q_\mathbf{k})_0}, 
  \label{eq:variationsQ}
\end{equation}
\begin{equation}
  \delta V(E_{\mathbf{k}})=\frac{V(E_\mathbf{k})_g-V(E_\mathbf{k})_0}{V(E_\mathbf{k})_0}, 
 \label{eq:variationsE}
\end{equation}
where 
\begin{equation}
V(X_\mathbf{k})_{g/0}=\left<\hat{X}_\mathbf{k}\hat{X}^\dagger_\mathbf{k}\right>_{g/0}-\left<\hat{X}_\mathbf{k}\right>_{g/0}\left<\hat{X}^\dagger_\mathbf{k}\right>_{g/0}
\end{equation}
is the variance of the field $X_\mathbf{k}=\{Q_\mathbf{k},E_\mathbf{k}\}$, $\hat{Q}_{\mathbf{k}}=\hat{b}_{\mathbf{k}} + \hat{b}_{-\mathbf{k}}^\dagger$, $\hat{E}_{\mathbf{k}}=\hat{a}_{\mathbf{k}} + \hat{a}_{-\mathbf{k}}^\dagger$ and $\left< ...\right>_g$ denotes the expectation value for a finite coupling $g$ while $\left< ...\right>_0$ denotes the expectation value for $g=0$. 

To compute the observables defined in Eqs.~(\ref{eq:variationsQ})-(\ref{eq:variationsE}), we use the open-system Truncated Wigner Approximation (TWA)~\cite{polkovnikov_phase_2010}. This is a phase-space semiclassical method that captures the first quantum correction to the mean-field approximation, and is adapted to include dissipation. This serves as an alternative to other approaches such as the master equation formalism, where increasing the Hilbert space, by increasing the number of modes $N$, hinders its numerical simulation. In TWA, instead of studying the density matrix operator dynamics, one formulates the problem using the Wigner quasiprobability distribution. The equations of motion for the Wigner distribution are truncated and solved by following the classical dynamics of many different realizations (trajectories) which are initially sampled over complex fields as dictated by the Wigner distribution of the uncoupled system. Each of these realizations follows an ordinary stochastic differential equation, the so-called Heisenberg-Langevin equations of motion, where dissipation and noise are introduced satisfying the fluctuation-dissipation theorem.

The Heisenberg-Langevin equations of motion for the complex cavity field $a_\mathbf{k}$ and complex Raman field $b_\mathbf{q}$ are
\begin{align}
\begin{split}
i\partial_{t}{a_\mathbf{k}} =& \omega_\mathbf{k}^c a_\mathbf{k} 
+ \frac{2g}{\sqrt{N}} \sum_{\mathbf{q}} (b_{\mathbf{k}-\mathbf{q}}+b_{-\mathbf{k}+\mathbf{q}}^{*})(a_\mathbf{q}+a_{-\mathbf{q}}^{*}) \\
&+ \frac{g_4}{N} \sum_{\mathbf{k'},\mathbf{q}}
   (a_{\mathbf{k}+\mathbf{q}}+a_{-\mathbf{k}-\mathbf{q}}^{*})
   (a_{\mathbf{k'}}+a_{-\mathbf{k'}}^{*}) \\
&\quad \times (a_{-\mathbf{k'}-\mathbf{q}}+a_{\mathbf{k'}+\mathbf{q}}^{*})
- i\kappa a_\mathbf{k} + i\xi_{a_\mathbf{k}},
\end{split}\label{eq:motion1} \\
\begin{split}
i\partial_{t}{b_\mathbf{q}} =& \omega_\mathbf{q}^R b_\mathbf{q} 
+ \frac{g}{\sqrt{N}} \sum_{\mathbf{k}}
  (a_\mathbf{k}+a_{-\mathbf{k}}^{*})(a_{-\mathbf{k}+\mathbf{q}}+a_{\mathbf{k}-\mathbf{q}}^{*}) \\
&- \tfrac{i\gamma}{2}(b_\mathbf{q}-b_\mathbf{q}^{*}) - \xi_{b_\mathbf{q}}.
\end{split} \label{eq:motion2}
\end{align}
Here $\kappa$, $\gamma$ are the decay rates associated with the cavity and Raman fields while $\xi_{a_\mathbf{k}}$ and $\xi_{b_{\mathbf{q}}}$ are sources of Gaussian noise obeying the autocorrelation relations $\left<\xi_{a_\mathbf{k}
}^{*}(t_1)\xi_{a_\mathbf{k}^\prime}(t_2)\right>=\kappa\delta(t_1-t_2)\delta_{\mathbf{k},\mathbf{k^\prime}}$ and $\left<\xi_{b_{\mathbf{q}}}(t_1)\xi_{b_{\mathbf{q^\prime}}}(t_2)\right>=\gamma \delta(t_1-t_2)\delta_{\mathbf{q},\mathbf{q^\prime}}$.
$\xi_{a_\mathbf{k}}$ is complex valued, whereas $\xi_{b_\mathbf{q}}$ is real valued. The last condition models that the dissipative force acting on the Raman modes is proportional to the imaginary part of the Raman field only, analogous to the Brownian motion problem, where the dissipative force is proportional to the velocity. The observables are computed in the steady state that is reached after a slow ramping of the Raman-cavity coupling from zero to $\frac{g}{\sqrt{N}}$ and the quartic nonlinear coupling from zero to $\frac{g_4}{N}$. We compute the expectation values by averaging over $3.5 \times 10^3$ realizations. We consider vacuum and thermal initial states with $\left<\hat{Q}_\mathbf{k}\right>_0=\left<\hat{E}_\mathbf{k}\right>_0=0$ and study the variation in the fluctuations.


We further simplify our analysis to consider a one-dimensional problem. We take $\mathbf{k}=(k_x,0,0)$ and $\mathbf{q}=(q_x,0,0)$, that we simply call $k$ and $q$ hereafter, and consider $N=11$ Raman modes and $N=11$ cavity modes with different energy dispersions. These modes correspond to $N=11$ different momenta: five positive momenta, five negative momenta, and the zero-momentum mode. Using the definitions of $\hat{Q}_{k}$ and $\hat{E}_{k}$ it is straightforward to show that $V(X_{k})=V(X_{-k})$ with $X_{k}=\{Q_{k},E_{k}\}$, so we only present the Raman and cavity fluctuations in the six non-negative momenta ($k\ge0$). 
TWA is expected to be accurate especially in the multimode scenario that we present here in the same way it is accurate in describing collective spin systems or systems with many matter degrees of freedom (emitters)~\cite{polkovnikov_phase_2010,Rabl2021,Rosario2025}.

\section{Numerical Results}

First we discuss the case in which all Raman active modes frequencies are degenerate and equal to $\omega_0^R$ and the cavity modes frequencies are also degenerate and equal to the photonic bandgap $\omega_0^c$ which we consider to be tunable. This corresponds to considering perfectly flat phonon and flat photon dispersions.

\begin{figure}[!tb]
\hspace*{-0.4cm} 
\includegraphics[width=0.45\textwidth]{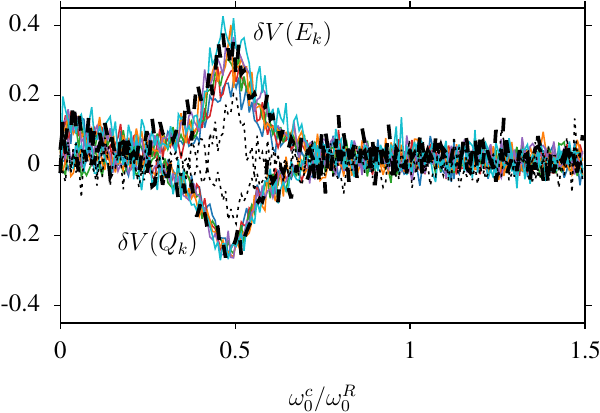}
\caption{Modification of the Raman and cavity field variances in each mode (different colors) as a function of the photonic band gap $\omega_0^c$. Amplification of photonic fluctuations ($\delta V(E_k)>0$) and attenuation of Raman fluctuations ($\delta V(Q_k)<0$) occur for $\omega_0^c\approx\omega_0^R/2$. Both, the phonon and photon bands are considered perfectly flat. We use $g=0.04\omega_0^R$, $\kappa=\gamma=0.02\omega_0^R$,  $g_4=0.01\omega_0^R$ and $N=11$. The black lines show the results for a single-mode case with $\frac{g}{\sqrt{N}}$ coupling (thin dashed line) and with $\sqrt{N}$-times larger effective coupling $g_{\mathrm{eff}}=g$ (wide dashed line).
} 
\label{fig:amplification}
\end{figure}

In Fig.~\ref{fig:amplification}, we plot the modification of the variance of the cavity field $\delta V(E_k)$ and the modification of the variance of the Raman coordinate $\delta V(Q_k)$ as a function of the photonic band gap $\omega_0^c$ in the six different Raman and cavity modes $k$ (different colors). In this case, where both dispersions are completely flat, the magnitude of the fluctuations in each mode is approximately the same, so the curves of different colors overlap.
At $\omega_0^c\approx\omega_0^R/2$ a parametric fluctuation amplification occurs. In this regime, the variance of the cavity field increases ($\delta V(E_k)>0$) and the variance of the Raman coordinate decreases ($\delta V(Q_k)<0$) in each mode with respect to the variances without coupling. Out of resonance $\delta V(E_k)\approx\delta V(Q_k)\approx0$ indicating that the cavity and Raman fluctuations are not modified despite hybridization. The parametric amplification of the equilibrium fluctuations was first reported in Ref.~\cite{HP24} for the single-mode scenario of Raman-cavity hybrids, which we also plot here for reference (thin black dashed lines). In the multimode case, the amplification of the cavity fluctuations and the attenuation of the Raman fluctuations, in each $k$ mode, are further enhanced owing to the collective many-body interactions [Note that the colored curves lie above the thin dashed lines]. Interestingly, we find that this many-body interaction-based enhancement can be captured by an effective single-cavity and single-Raman mode model, considering an effective Raman-light coupling $\sqrt{N}$ times larger than $\frac{g}{\sqrt{N}}$ $(g_{\mathrm{eff}}=g)$. The variation in fluctuations using this effective model is plotted with wide dashed lines and overlaid on the many-body simulation, shown in colors. Although the total variances of the cavity field increase in resonance and overcome the variance in the vacuum state $V(E_k)_{0}=1/4$, the cavity field is relatively squeezed in the new ground state of the interacting system (see Appendix A). Additionally, not only do the Raman fluctuations change due to hybridization with the cavity but also the Raman coordinate at zero momentum does $(\left<\hat{Q}_{\mathbf{k}=0}\right>_g\ne 0)$ (see Appendix B). Notably, both the shift in the Raman coordinate and cavity squeezing become larger in the multi-mode scenario.

At this point, it is important to emphasize that the hybrid system is in equilibrium and no coherent driving force is considered. Interestingly, multimode Raman fluctuations (noise) act as such and produce an amplification of the cavity fluctuations in each mode for $\omega_0^c\approx\omega_0^R/2$. This constitutes an extension of the concept of parametric amplification in driven systems to the case of zero-point energy fluctuations~\cite{HP24}. This phenomenon becomes even more evident for \emph{multimode} Raman-cavity hybrids, as illustrated in Fig.~\ref{fig:amplification}.


\begin{figure}[!tb]
\hspace*{-0.0cm} 
\includegraphics[width=0.45\textwidth]{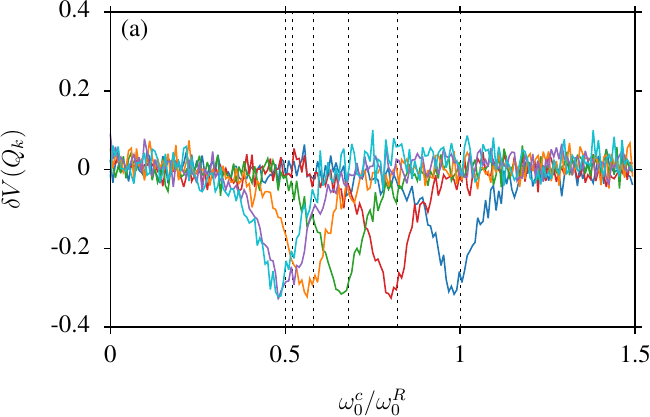}
\includegraphics[width=0.45\textwidth]{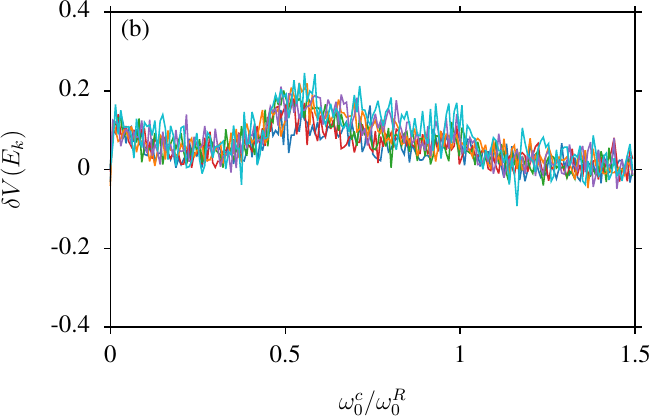}
\caption{Modification of the variance of the Raman fluctuations $\delta V(Q_k)$ (a) and variance of the photonic fluctuations $\delta V(E_k)$ (b) as a function of the photonic band gap $\omega_c^0$. The different colors indicate the six different modes with $k\ge0$ (see text). We consider a perfectly flat photon dispersion whereas the Raman dispersion is considered to be quadratic in $k$ with a bandwidth $\delta\omega_k^R=\omega_0^R$. The vertical dashed lines in (a) indicate the resonant condition $2\omega_0^c=\omega_k^R$ for the different Raman modes $k$. The rest of parameters are as in Fig.~\ref{fig:amplification}.
} 
\label{fig:Ramandispersion}
\end{figure} 

We now present two examples where either the phonon or the photon band is not flat but quadratic in $k$. In Fig.~\ref{fig:Ramandispersion} we show the Raman and cavity fluctuations for a flat photonic band and a quadratic Raman dispersion $\omega_k^R=\omega_0^R(1+k^2/k^2_{max})$ so the Raman bandwidth is $\delta\omega_k^R=\omega_{k_{max}}^R-\omega_0^R=\omega_0^R$. We plot the variance modification of the fluctuations in the six nonnegative $k$ modes in the interval $k\in[0,k_{max}]$. In this case, Raman fluctuations in each mode can be selectively reduced by tuning the photonic bandgap. The Raman mode $k$ with $\omega_k^R\approx2\omega_0^c$ is the one at which the fluctuations are strongly attenuated, while the other modes are less affected [Fig.~\ref{fig:Ramandispersion} (a)]. For example, by setting the photonic bandgap to $\omega_0^c\approx\omega_0^R$, only the fluctuations in the highest momentum Raman mode $k_{max}$ (dark-blue curve) are significantly attenuated, so that $\delta V(Q_{k_{max}})\approx-0.3$. The fluctuations in the other Raman modes do not change, for this photonic bandgap, and $\delta V(Q_k)\approx0$. This level of selectivity is higher for large momenta, since for these modes the resonance width, which is determined by the decay rates, is smaller than the energy difference between two consecutive Raman modes.

\begin{figure}[!tb]
\hspace*{-0.4cm} 
\includegraphics[width=0.45\textwidth]{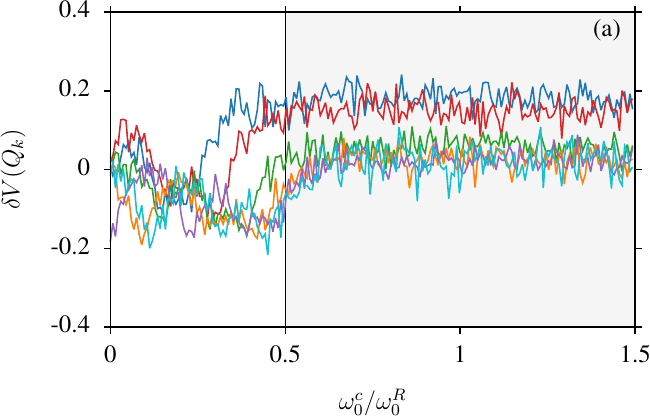}
\hspace*{-0.4cm} 
\includegraphics[width=0.45\textwidth]{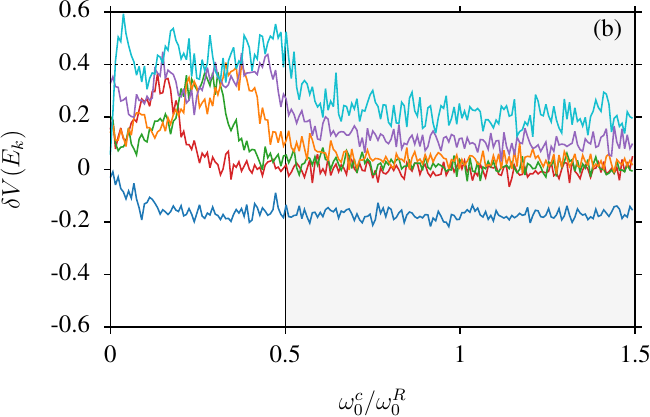}
\caption{Same as Fig.~\ref{fig:Ramandispersion} but considering a flat Raman dispersion and a quadratic cavity dispersion with a bandwidth $\delta\omega_c^k=\omega_0^R$ (see main text). The vertical dashed lines at $\omega_0^c=0.5\omega_0^R$ indicate the threshold above which resonances are not possible (gray zones). The horizontal dashed line in the lower panel (b) marks the maximum value of $\delta V(E_k)$ that can be obtained by considering a collective enhancement factor of $\sqrt{N}$ on the coupling, i.e. the maximum cavity fluctuation value reached in Fig.~\ref{fig:amplification} for perfectly flat bands. Notice the different scale in the vertical axis in the lower panel compared to the vertical axes of all previous figures.
} 
\label{fig:Cavitydispersion}
\end{figure}

Fig.~\ref{fig:Ramandispersion} (b) shows the modification of the variance of the cavity field in each mode $\delta V(E_k)$. It exhibits a weak and broad resonance around $\omega_0^R/2<\omega_0^c<\omega_0^R$ meaning that the cavity fluctuations are slightly amplified in this non-flat Raman-dispersion case. The fluctuations change in the same amount in each mode. However, photonic fluctuations are less sensitive to the presence of Raman fluctuations and $|\delta V(E_k)|<|\delta V(Q_k)|$ when in resonance. It suggests that Raman fluctuations in the different non-degenerate Raman modes are less effective at amplifying the cavity fluctuations. Therefore there is a non-reciprocal redistribution of fluctuations between the cavity and Raman modes.
Overall, this result demonstrates how, by tuning the photonic bandgap, it is possible to selectively and non-reciprocally control phonon fluctuations in dispersive Raman materials, which could have potential applications in controlling the properties of matter through cavities.

In Fig.~\ref{fig:Cavitydispersion} we show the same observables but now considering a flat Raman band with frequency $\omega_0^R$ and a quadratic cavity dispersion $\omega_k^c=\omega_0^c(1+\frac{\omega_0^Rk^2}{\omega_0^ck^2_{max}})$ with $k\in[0,k_{max}]$. 
In this case, we identify two distinct regimes. If $\omega_0^c > \omega_0^R/2$ (gray shaded region), $\omega_k^c > \omega_k^R/2$ for all $k$, and there is no resonant modification of the fluctuations, since it is not possible to convert two photon mode frequencies to the Raman mode frequency $\omega_0^R$ while conserving energy. This is why, in this regime, there are no peaks in $\delta V(E_k)$ or $\delta V(Q_k)$ for any $k$. Still, owing to the multimode interactions, the fluctuations are nonresonantly modified and $\delta V(E_k)$ and $\delta V(Q_k)$ are nonzero in general. Remarkably, for high-frequency modes, the Raman fluctuations are amplified instead of being attenuated. This is in marked contrast to the single-mode case, where the Raman-mode fluctuations either decrease at resonance or remain constant out of resonance, but do not increase. This phenomenon is illustrated in Fig.~\ref{fig:Cavitydispersion} (a) where $\delta V(Q_k)>0$ in the non-resonant regime $\omega_0^c>\omega_0^R/2$ for the two highest momentum modes (blue and red curves). A similar result is obtained for the fluctuations of the cavity field in the $k_{max}$ mode [dark-blue curve in Fig.~\ref{fig:Cavitydispersion} (b)]. In this case, the cavity fluctuations are nonresonantly attenuated instead of amplified so $\delta V(E_{k_{max}})<0$.

For $\omega_0^c<\omega_0^R/2$ resonances are possible and several peaks in the deviation of the variances $\delta V(Q_k)$ and $\delta V(E_k)$ are visible in Fig.~\ref{fig:Cavitydispersion} (a) and Fig.~\ref{fig:Cavitydispersion} (b), respectively. 
Since the cavity frequencies are nondegenerate, there are several ways in which two different cavity modes with frequencies $\omega_k^c$ and $\omega_{k^\prime}^c$ convert into the Raman frequency $\omega_0^R$. This leads to a richer redistribution of Raman and cavity fluctuations in each mode and the appearance of several peaks in $\delta V(Q_k)$ and $\delta V(E_k)$ by tuning the photonic band gap. Here, there is also a non-reciprocal effect due to the different phonon and photon dispersions. The cavity field fluctuations are more efficiently modified by the Raman fluctuations, while these also vary, but on a smaller scale. In the resonant regime with $\omega_0^c<\omega_0^R/2$, for most $k$ modes, $|\delta V(E_k)|>|\delta V(Q_k)|$.

Interestingly, the $\sqrt{N}$ enhancement in fluctuations is exceeded in certain cavity modes when non-flat bands are considered. The modification of the fluctuations is larger than that obtained using an effective model of a single cavity mode coupled to a single Raman mode with a $\sqrt{N}$ times larger coupling constant. This is the case for cavity fluctuations in the zero-momentum mode. $\delta V(E_0)$, shown with a cyan curve in Fig.~\ref{fig:Cavitydispersion} (b), is larger than the maximum deviation of cavity fluctuations obtained for perfectly flat bands $\delta V(E_k)_{max}=0.4$ (c.f. Fig.~\ref{fig:amplification}). This suggests that it is possible to significantly enhance the fluctuations in specific modes by engineering the photon dispersion.

We now extend our analysis to thermal fluctuations. We consider that the hybrid multimode system is coupled to a common thermal bath which is equilibrated at a certain temperature $T$, so the photon and phonon population is finite and the Wigner distributions associated with each mode are broader. In this regime, the possible negativity of the Wigner distribution, due to quantum effects, vanishes and semiclassical methods such as TWA are very accurate.

\begin{figure}[!tb]
\includegraphics[width=0.45\textwidth]{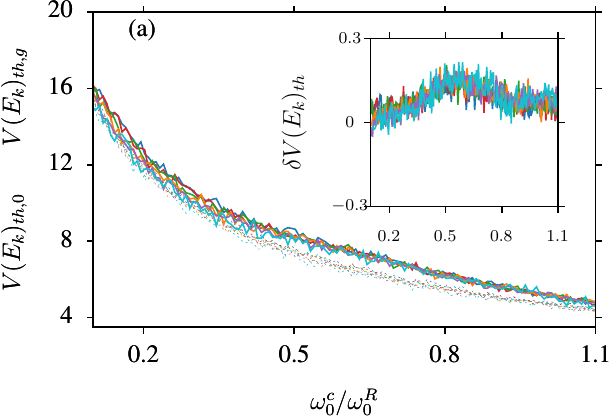}
\includegraphics[width=0.45\textwidth]{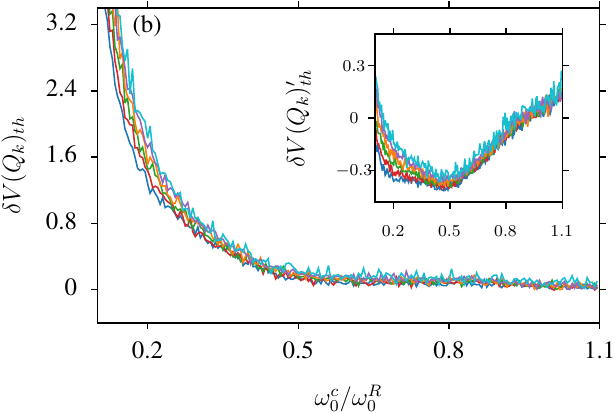}
\caption{Thermal cavity and Raman fluctuations. (a) Variance of the cavity field in each $k$ mode (different colors) in the presence of multiple Raman modes (solid lines). The variance of the cavity field in the absence of Raman fluctuations is shown with thin dotted lines for reference. The difference between these variances is shown in the inset. (b) Change in the variance of the Raman coordinate, in each $k$ mode, with respect to the variance of the Raman coordinate in the absence of cavity fluctuations. The difference of the variance of the Raman coordinate with respect to the variance of the cavity field in the interacting regime is shown in the inset (see text). The phonon and photon bands are considered flat and $\hbar \omega_0^R=
0.5k_B T.$ The rest of the parameters are as in Fig.~\ref{fig:amplification}} 
\label{fig:thermal}
\end{figure}

In Fig.~\ref{fig:thermal} we show how the thermal fluctuations of the cavity and the Raman fields change as a function of the photonic band gap by considering flat photon and phonon dispersions and a fixed temperature. Specifically, we consider $\hbar \omega_0^R=0.5k_B T$ so the phonon occupation number is $\sim2$ and the number of photons vary between $\sim2$ and $\sim20$ for the different photonic band gaps that we explore.

In the absence of Raman fluctuations and keeping the bath temperature constant, the variance of the free cavity fluctuations increases as the photonic band gap decreases as $\sim\frac{1}{2}\coth(\frac{\hbar \omega^c_k}{2 k_B T})$.
The variance of such free cavity fluctuations in each mode, $V(E_k)_{th,0}$, is plotted in Fig.~\ref{fig:thermal} (a) with dotted colored lines for reference. Those thermal fluctuations are altered by the coupling to the thermal Raman-mode fluctuations. The variance of the cavity field, in each mode $k$, for the coupled system, $V(E_k)_{th,g}$, is plotted with full colored lines. The difference of those variances, properly renormalized, $\delta V(E_k)_{th}=\frac{V(E_k)_{th,g}-V(E_k)_{th,0}}{V(E_k)_{th,0}}$, is shown in the inset. This illustrates how the thermal cavity fluctuations are also amplified by the thermal Raman fluctuations around the resonant condition $\omega_0^c\approx\omega_0^R/2$. 

The modification of the variance of thermal Raman fluctuations $\delta V(Q_k)_{th}=\frac{V(Q_k)_{th,g}-V(Q_k)_{th,0}}{V(Q_k)_{th,0}}$ due to the coupling to the cavity is shown in Fig.~\ref{fig:thermal} (b). 
The Raman fluctuations in each mode $V(Q_k)_{th,g}$, and consequently $\delta V(Q_k)_{th}$, increase with decreasing cavity bandgap $\omega^c_0$. This increase is dominated by the increase in cavity thermal fluctuations to which the Raman modes are coupled. Notice that in the uncoupled situation, the variance of Raman fluctuations in free space, $V(Q_k)_{th,0}\sim\frac{1}{2}\coth(\frac{\hbar \omega^R_k}{2 k_B T})$, remains constant regardless of the photonic bandgap value. One question one can ask is how this increase in Raman fluctuations for small photonic bandgap compares to the increase of cavity fluctuations shown in Fig.~\ref{fig:thermal} (a). In the inset of Fig.~\ref{fig:thermal} (b), we show the modification of the variance of thermal Raman fluctuations with respect to the variance of the cavity field fluctuations $\delta V(Q_k)_{th}^\prime=\frac{V(Q_k)_{th,g}-V(E_k)_{th,g}}{V(E_k)_{th,g}}$ instead of computing it with respect to the variance of thermal Raman fluctuations in free space as previously considered [$\delta V(Q_k)_{th}$].
$\delta V(Q_k)_{th}^\prime$ exhibits a reduction around $\omega_0^c\approx\omega_0^R/2$ indicating a relative decrease of the Raman fluctuations similarly to what we discussed in the previous cases in the quantum noise limit (zero-temperature regime).
Overall, this suggests that the collective fluctuation amplification engineering previously demonstrated for flat bands is also feasible for thermal states.

\section{Experimental Realizations}

In this section, we briefly discuss possible platforms for realizing multimode Raman-cavity hybrids.

In general, Raman-active two-dimensional (2D) materials are good candidates for realizing our proposal. 2D materials are versatile and offer high tunability to be integrated in different cavity settings; ranging from split ring resonators, Fabry-Perot cavities, surface phonon polaritons and photonic crystals in the THz range. Transition-metal dichalcogenides (TMDs) are particularly interesting, as THz phonon modes can be tuned with the number of layers and the twist angle~\cite{Alex2016,Lin2021}.

We choose a photonic crystal for the derivation of our model in Section II but other types of cavity-QED settings can also be used to implement our proposal. Photonic crystals have been widely studied in the last decades and the possibility to engineer their band dispersions in the THz range~\cite{Beigang} makes them a promising candidate for exploring nontrivial collective fluctuations in the many-body scenario. Alternatively, phonon-polaritons-based platforms playing the role of photons and Fabry-Perot cavities with Raman active solid-state materials also align with our model~\cite{HP24}.

Another interesting realization of our model is the implementation via dark excitons in TMDs where a parametric nonlinear coupling between two photons and one dark exciton has been experimentally demonstrated~\cite{Viviana24}. The dispersive dark exciton band could couple to an inherent or integrated multimode cavity giving rise to a phenomenology similar to what we have presented in this paper. Raman spectroscopy of such light-matter hybrids could be used to characterize the possible collective amplification phenomena and nontrivial fluctuations in each mode based on the different Rabi splitting in the spectrum~\cite{HP24}.

Ions traps offer a different platform in which the nonlinear bosonic interactions as in the third term of Eq.(\ref{HamiltonianMB}) can be simulated. In this case, instead of photon-phonon interactions, the nonlinear parametric coupling can be realized between different transverse and longitudinal phonon modes via the presence of Coulomb interaction between the ions. Here, the transverse phonon modes could play the role of cavity modes and the longitudinal phonon modes could play the role of Raman modes. A first experimental realization of such a parametric coupling has been demonstrated using two trapped ions in Ref.~\cite{trappedions}. This constitutes a phonon-based simulator of the single-mode Raman-cavity case. Following Ref.~\cite{multipletrappedions}, an extension to a multimode scenario is also envisaged using a linear ion trap with many ions. Depending on the number of confined ions and the anisotropy of the trapping potential, it is possible to have multiple transverse phonon modes acting on a second longitudinal phonon mode in a parametric fashion, as in the third term of Eq.~(\ref{HamiltonianMB}). In this platform, the different phonon modes (longitudinal and transverse) are associated with the different ions rather than different momenta, so although the nonlinear coupling between different modes is of the form $\sim g_{lmn}(\hat{a}_l+ \hat{a}_l^\dagger)(\hat{a}_m+ \hat{a}_m^\dagger)(\hat{b}_n+ \hat{b}_n^\dagger)$, the indexing indicating which mode interacts with which mode is different. For example, in this ion trap realization, $g_{lmn}\neq 0$ only if at least two indices are identical. For further details, we refer the readers to Ref.~\cite{multipletrappedions} and specifically to the discussion around Eq. (34) and the tables shown in the appendix.


An advantage of this ion trap setup is the direct access to the Wigner function distribution and thus the direct access to transverse phonon fluctuations without the need for additional spectroscopic techniques~\cite{trappedions}. This platform, which shares key similarities with our system, constitutes a rich platform in its own right. The possibility of detecting similar phenomena to those discussed in this paper and studying the parametric manipulation of phonon fluctuations would be presented elsewhere.


Finally, molecular polaritons provide a mature platform; see~\cite{Herrera_Owrutsky2020,SanchezBarquilla_Feist2022,Xiang2024}, in which parametric amplification engineering can be explored. Relevant scenarios include non-resonant spontaneous Raman scattering in molecules~\cite{Galland2023}, as well as emergent nonlinear effects~\cite{Yuen-Zhou2025} that can be investigated in cavity configurations and off-resonant surface-enhanced Raman spectroscopy setups~\cite{Galland2016,Aizpurua2016,Hughes2017}, beyond the standard optomechanical regime (where squeezed-photon contributions of the form $\hat{a}\hat{a}$ and $\hat{a}^\dagger\hat{a}^\dagger$ become significant).

\section{Conclusions}
In this paper, we have generalized single-mode Raman-cavity hybrids; see~\cite{HP24,Keeling2024}, to a system that has nonlocal multimode Raman-cavity interactions.
For this multimode scenario, we demonstrate how to advance the resonant control of Raman and cavity fluctuations in the single-mode model in several ways by engineering the photon and phonon dispersions. For flat phonon and photon dispersions, the Raman coupling further enhances the amplification and localization of cavity and Raman fluctuations in each mode, respectively. We demonstrate that this collective amplification increases with the number of interacting Raman and cavity modes as $\sqrt{N}$. For not-flat bands, we demonstrate: (i) selective and nonreciprocal control of Raman fluctuations by the cavity [Fig.~\ref{fig:Ramandispersion}], (ii) cavity fluctuation amplification exceeding $\sqrt{N}$ scaling in specific modes, and, in notable contrast to the single-mode scenario, (iii) non-resonant control of Raman and cavity fluctuations that can be amplified or attenuated at will by choosing the appropriate mode and band gap $\omega_0^c$ [see Fig.~\ref{fig:Cavitydispersion}].
The degree of control offered by dispersion engineering over the fluctuations in these nonlinear cavity-matter hybrids makes this platform promising not only for cavity control of quantum materials by noise manipulation but also for optimizing quantum sensing and spectroscopy.

\section*{Acknowledgments} 
We acknowledge funding by the Deutsche Forschungsgemeinschaft (DFG, German Research Foundation) “SFB-925” Project No 170620586 and the Cluster of Excellence “Advanced Imaging of Matter” (EXC 2056), Project No. 390715994.

\appendix

\section{Squeezing of the cavity field}

\begin{figure}[!tb]
\includegraphics[width=0.45\textwidth]{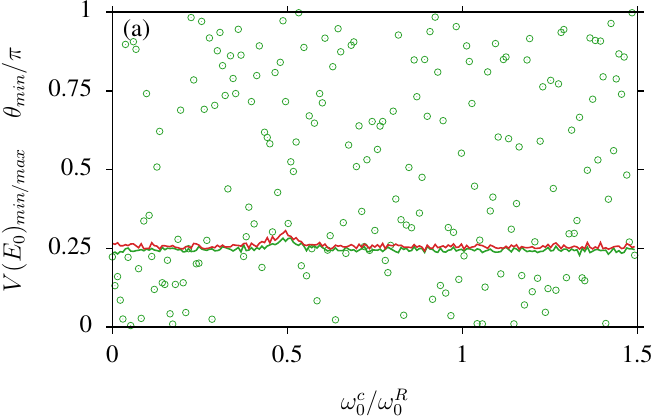}
\includegraphics[width=0.45\textwidth]{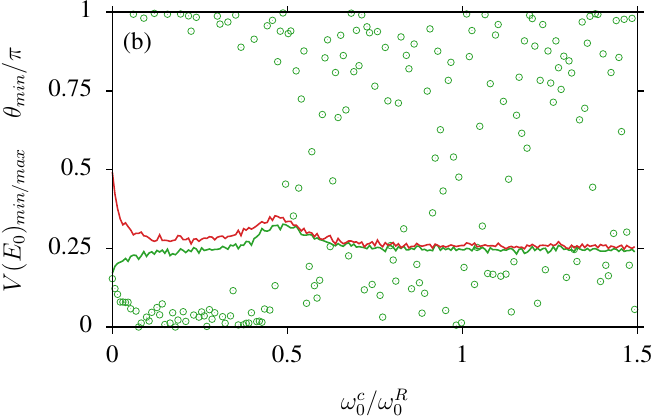}
\caption{Squeezing of the cavity field in the zero-momentum mode. Minimum (green) and maximum (red) variance of the cavity field in the zero-momentum mode for the single (a) and multimode (b) case. The angle at which the variance is minimum, $\theta_{min}$ is shown with empty green circles in each case. The rest of the parameters are as in Fig.~\ref{fig:amplification}.}. 
\label{fig:squeezing}
\end{figure} 

\begin{figure}[!tb]
\includegraphics[width=0.45\textwidth]{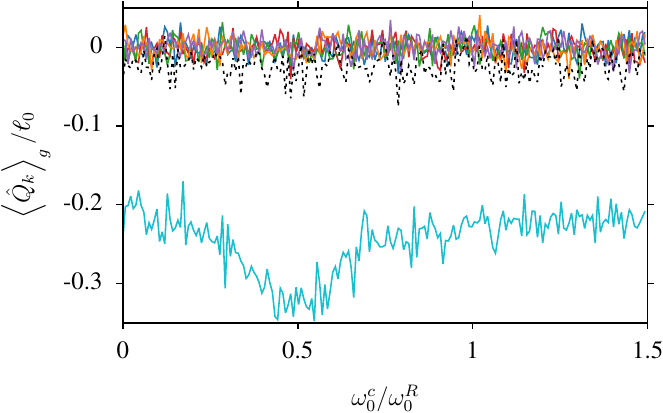}
\caption{Raman coordinate as a function of the photonic band gap in the different momenta modes (different colors) for the many-body case. The Raman shift for the single-mode case is plotted with a black dashed line as a reference. In the multimode scenario the Raman coordinate shift only in the zero-momentum mode (cyan curve). The Raman coordinate is plotted in the natural unit length $\mathcal{l}_0=\sqrt{\hbar/M\omega_0^R}$. The rest of the parameters are as in Fig.~\ref{fig:amplification}.}. 
\label{fig:Ramanshift}
\end{figure}

In Fig.~\ref{fig:squeezing} we show the minimum and maximum variance of the cavity field in the zero-momentum mode as a function of the photonic band gap, for the single and multimode cases. Additionally, we also show with empty dots the angle $\theta_{min}$ at which the variance is minimum. The maximum variance occurs at $\theta_{min}\pm\pi/2$.
In panel (a), the variances increase at the $\omega_0^c=0.5\omega_0^R$ resonance and the maximum and minimum variances are quite close, so the relative squeezing of the ground state of the hybrid system is relatively small in this single-mode case. In the many-body scenario [Fig.~\ref{fig:squeezing} (b)], the difference between the maximum and minimum variance is more noticeable, indicating an increase in the relative squeezing of the cavity field due to interactions. The difference between the minimum and maximum variances in the different non-zero momentum modes [not shown] are less affected by the multi-mode coupling and remain similar to the single-mode case, where the squeezing is lower. 

In the single-mode case the Wigner distribution of the cavity field is almost not squeezed and very much round in the phase-space, so the angle at which the minimum variance is found is practically random. In the multimode scenario however the squeezing notably increases and the minimum variance appears at the two equivalent angle $\theta_{min}\simeq0$ or $\theta_{min}\simeq\pi$.

\section{Raman shift}
In Fig~\ref{fig:Ramanshift} we show the expectation value of the Raman coordinate in the different modes as a function of the photonic bandgap. Due to many-body interactions, the Raman coordinate in the zero-momentum mode is further shifted relative to the shift obtained in the single-mode case with $\frac{g}{\sqrt{N}}$ coupling (thin dashed line). The Raman coordinate remains at its equilibrium value in the other non-zero-momentum modes. The zero-momentum Raman shift occurs for any value of the photonic band gap although a resonant enhancement is observed in the multimode scenario at $\omega_0^c\approx\omega_0^R/2$ (cyan curve).

\bibliography{references}

\end{document}